# Development of a timing chip prototype in 110 nm CMOS technology

M Senger[1], L Caminada[1,2], B Kilminster[1], A Macchiolo[1], B Meier[2] and S Wiederkehr[1,2]

[1]Universität Zürich, Physik-Institut, Winterthurerstrasse 190, CH-8057, Zurich, Switzerland
[2]Paul Scherrer Institut, Forschungsstrasse 111, 5232 Villigen PSI, Switzerland

E-mail: matias.senger@cern.ch

25 May 2021

**Abstract.** We present a readout chip prototype for future pixel detectors with timing capabilities. The prototype is intended for characterizing 4D pixel arrays with a pixel size of $100 \times 100$ µm$^2$, where the sensors are Low Gain Avalanche Diodes (LGADs). The long-term focus is towards a possible replacement of disks in the extended forward pixel system (TEPX) of the CMS experiment during the High Luminosity LHC (HL-LHC). The requirements for this ASIC are the incorporation of a Time to Digital Converter (TDC) within each pixel, low power consumption, and radiation tolerance up to $5 \times 10^{15}$ $n_{\text{eq}}$ cm$^{-2}$ to withstand the radiation levels in the innermost detector modules for 3000 fb$^{-1}$ of the HL-LHC (in the TEPX). A prototype has been designed and produced in 110 nm CMOS technology at LFoundry and UMC with different versions of TDC structures, together with a front end circuitry to interface with the sensors. The design of the TDC will be discussed, with the test set-up for the measurements, and the first results comparing the performance of the different structures.

## 1. Introduction

Accelerator-based high energy physics (HEP) experiments will collect data at substantially higher instantaneous and integrated luminosities over the next decades. This comes together with a number of challenges for the technology involved. For detectors, these higher luminosities translate into higher occupancy and higher radiation doses. While this applies to all detector components, the inner tracking systems of LHC experiments, which are nearest to the interaction region, are exposed to the highest data rates and irradiation. Higher pileup values translate into more difficulty in the reconstruction of events. Higher radiation levels translate into a faster degradation of the hardware in the detector.

The HL-LHC is the next phase of the LHC operation, beginning in 2027, that will increase the integrated luminosity by a factor of 10. The mean number of events per bunch crossing is expected to increase from 27 (design value) to 140-200 [1–3]. This is a factor between 5 and 7 higher than the design value for the detectors, such as ATLAS or CMS. Currently, the tracking system in these detectors provides only spacial information and with the higher pileup values expected it is challenging to maintain the current performance for vertex identification and heavy quark tagging. To overcome this, ATLAS and CMS are adding timing information in order to disentangle events that occur at the same position but at different times (within a bunch crossing) [2,3]. These timing layers in the CMS experiment will be done with scintillators in the central region and with LGAD silicon detectors in the forward region out to $|\eta| < 3$. The time resolution per hit is specified to be 30-70 ps. The LGADs in the forward region have an area of $\approx 3$ mm$^2$.

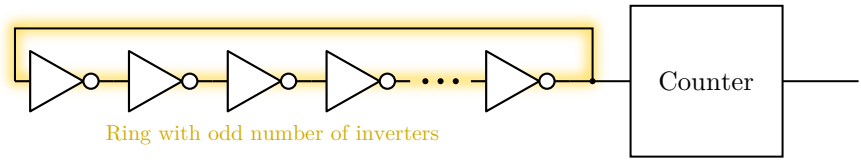

**Figure 1.** Basic concept for our TDC designs.

To make use of precision timing for $3 < |\eta| < 4$ in the CMS experiment, as an upgrade to the TEPX tracker pixel extension [4], we require smaller LGAD sensors with a pixel size of approximately $100 \times 100$ µm$^2$, and a time resolution of about 30 ps.

The development of LGAD sensors with a granularity similar to that of the present pixel detector is anyhow very challenging. An intense R&D effort is ongoing to obtain the required spacial resolution with this technology while maintaining the temporal resolution [5–10] in order to obtain the so called *4D pixels*, i.e. pixels capable of determining position and time of impact of a particle. The common belief in the community is that the bottleneck towards the 4D pixels is currently in the readout electronics rather than in the sensing technology, as was pointed out a number of times during the TIPP 2021 conference.

In this work we present the status of our R&D towards the incorporation of timing measurements in a readout ASIC for 4D pixels, with a goal of achieving a possible upgrade to the CMS TEPX detector. Here we focus on the TDC only. The design, implementation and performance measurements of two TDC variants are presented.

**2. TDC designs**

Two TDC designs were developed, implemented and tested. These are:

(i) Fully digital TDC with fast inverters and digital readout.
(ii) Semi-analog TDC with slower inverters and analog readout.

The two designs are based on the same concept as depicted in Figure 1. Here we see a closed path, i.e. a ring, with an odd number of inverters that comprises a ring oscillator. After a proper initialization a single "inverting wave" will travel along this ring and the counter will increase each time this wave passes through the last node, before it feeds back to the beginning of the ring. The oscillation period of the ring is given by the number of inverters and the propagation delay of each of them. After some time $\Delta t$ from the start of the oscillation, the time $\Delta t$ is measured by determining the position of the wave within the ring of inverters and the number of full loops registered by the counter. The two designs differ in the way in which $\Delta t$ is extracted.

In the following sections the two TDC designs will be presented.

*2.1. Fully digital TDC*

The fully digital TDC design makes use of the digital outputs from the inverters and counter. A simplified block diagram is shown in Figure 2. This design follows an architecture known as Vernier. It has two chains of identical fast inverters. One of these chains has 20 inverters and is closed through a NAND gate to which the START signal is connected. The other chain has 22 inverters and is not closed. There are also 21 differential flip-flops that sample each stage of these two chains of inverters, as shown. Finally, the last inverter in the auxiliary chain is connected to the input of a 7-bit counter.

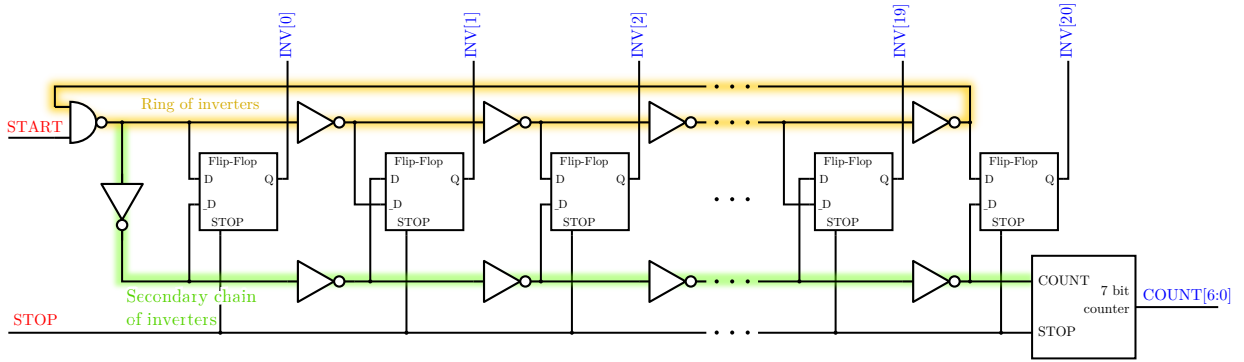

**Figure 2.** Fully digital TDC block diagram. Red signals (START and STOP) are inputs while blue signals (INV[:] and COUNT[:]) are outputs.

The principle of operation of this TDC is simple. Initially both START and STOP inputs are at a low level. The output of the NAND gate connected to the START signal is therefore a logic 1. This fixes the state of all the other nodes in both chains, alternating between 1 and 0. The counter is initialized to 0. If the outputs INV[:] and COUNT[:] were read at this point, we would obtain all 0's for the INV bus (due to the alternating connection of the flip-flops inputs, see Figure 2), and 0 for the counter. When the START signal changes to 1, the NAND gate becomes a NOT gate and allows the propagation of a "wave of inversions" along the two lines. After some time $\Delta t$ the STOP signal also changes to 1 and makes the flip-flops to sample the state of the chain and the counter to stop. The INV[:] bus will then be in a fixed state of the form, e.g., 000000000011111111111 meaning that the wave has had time to propagate through the first 11 inverters. We can thus map each output COUNT|INV (from now on this notation will be used to label the outputs of the TDCs) to some measured time, e.g. 2|000000000011111111111→1 ns. As we see, the INV part acts as a fractional division for the integer part given by COUNT.

This TDC was implemented in UMC 110 nm technology and it required an area of $110 \times 60$ µm$^2$.

*2.2. Semi-analog TDC*

The main conceptual difference between the semi-analog TDC design with respect to the fully digital is that the state of the chain of inverters is read as an analog signal. In this way it is possible to gain sub-digital resolution as there is always, for any $\Delta t$ between the START and the STOP signal, one of the inverters in the middle of a transition. Though the readout becomes more complex, the inverters now don't need to be as fast as in the fully digital design.

A simplified block diagram of the semi-analog design is shown in Figure 3. The basic principle is the same as was described for the fully digital TDC. This time, however, the inverters are isolated from the power supply when the STOP signal arrives, as illustrated in the diagram, in such a way that the amount of charge that was at each output is isolated, and can then be analogly measured through the INV lines. The parasitic capacitance at each input is used as a storage mechanism for this charge.

This TDC was implemented in LFoundry 110 nm technology and the required space was approximately $25 \times 50$ µm$^2$.

**3. Test setup**

A dedicated test setup was implemented to measure the performance of the TDCs. A simplified block diagram of it is shown in Figure 4. The entrance block is the Raspberry Pi which automates the measurement routines and stores the measured data, next there is an FPGA to interface with the hardware, followed by a specialized delay module, and finally the TDC under test. The diagram shows also an ADC block, which was only used when testing the semi-analog TDC. The delay module has two SY89296 delay chips and can be programmed to produce any $\Delta t$ between the START and STOP signals between -10 ns and 10 ns in steps of about 1 ps with fluctuations smaller than 5 ps [11, 12].

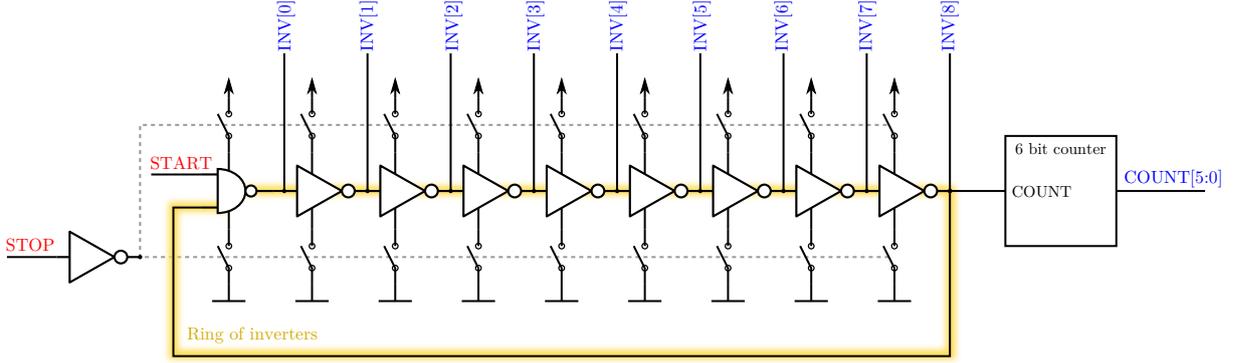

**Figure 3.** Semi analog TDC block diagram. Red signals (START and STOP) are inputs while blue signals (INV[:] and COUNT[:]) are outputs.

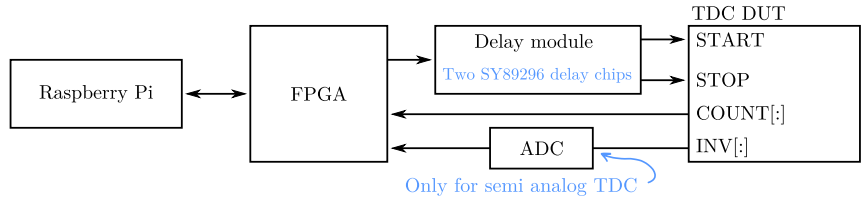

**Figure 4.** Block diagram of the test setup.

## 4. Results

Starting with the results of the fully digital TDC, in the left panel of Figure 5 the distribution of the first 7 outputs in $\Delta t = t_{\text{START}} - t_{\text{STOP}}$ is shown. A similar pattern, with different outputs, continues up to 10 ns which is the maximum tested $\Delta t$. To obtain a temporal resolution for the device, two quantities were studied: the standard deviation $\sigma$ and the width of each distribution $W$ measured as $W = q_{95\%} - q_{5\%}$ where $q_{x\%}$ is the $x\%$ quantile. While $\sigma$ is commonly used, $W$ is more representative for uniform distributions ($W$ is "the width of the distribution" and $\sigma \approx \frac{W}{\sqrt{12}}$ for approximately uniform distributions). The distribution of each of these two quantities, using all the data from 0 to 10 ns, is shown in the right panel of Figure 5. We can see that the width $W \lesssim 60$ ps, i.e. the "full width resolution in the range 0→10 ns" is better than 60 ps, and the standard deviation $\sigma \lesssim 20$ ps. The mean value for each quantity is $\approx 30$ ps and $\approx 10$ ps respectively.

The results of the semi-analog design are shown in the left panel of Figure 6, where a fragment of the raw signals measured out of the analog lines of the INV[:] bus as a function of $\Delta t$ is shown. Odd lines were inverted to ease the visualization, as denoted in the legend. It can be seen how each inverter changes its state smoothly between low and high levels. Also visible are the propagation delays from one inverter to the next one. Each line in this bus can be digitized with an arbitrary number of bits. If we use only one single bit we obtain results similar to those of the fully digital TDC, shown in the left panel of Figure 5, but with wider distributions as the inverters in this design are slower (so worse time resolution). We can, however, increase the number of bits for the digitization and gain analog resolution. In doing so it is possible to resolve the smooth transition between low and high states of the inverters and translate this into an improvement of time resolution. As an example, here we have chosen to show only the results with 3 bits, which are shown in the right panel of Figure 6 (the curious reader is directed to reference [13] where more details are given). In this plot we find the same quantities that were used for the fully digital TDC, i.e. the width $W$ and the standard deviation $\sigma$. As can be seen, the full width is $W \lesssim 30$ ps and the standard deviation is $\sigma \lesssim 10$ ps.

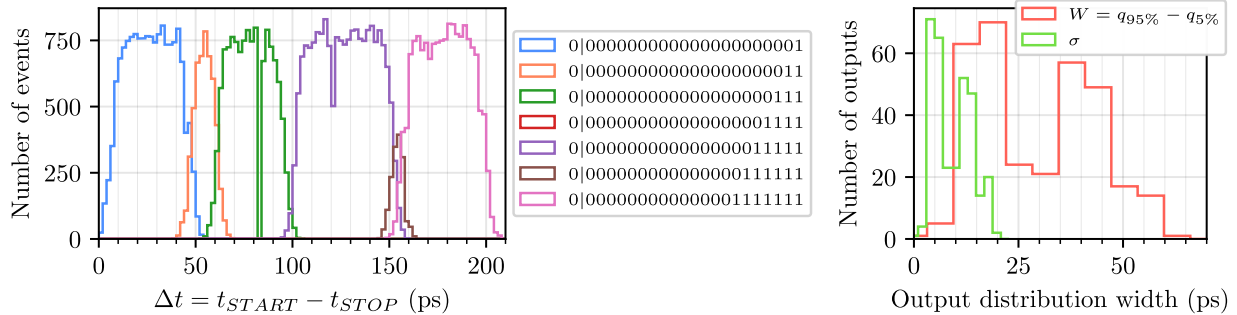

**Figure 5.** Left: Temporal distribution of the first seven outputs from the fully digital TDC. Right: Distribution of the time spread of each output (as the ones shown in the left panel) measured with the "full width" $W$ and with the standard deviation $\sigma$.

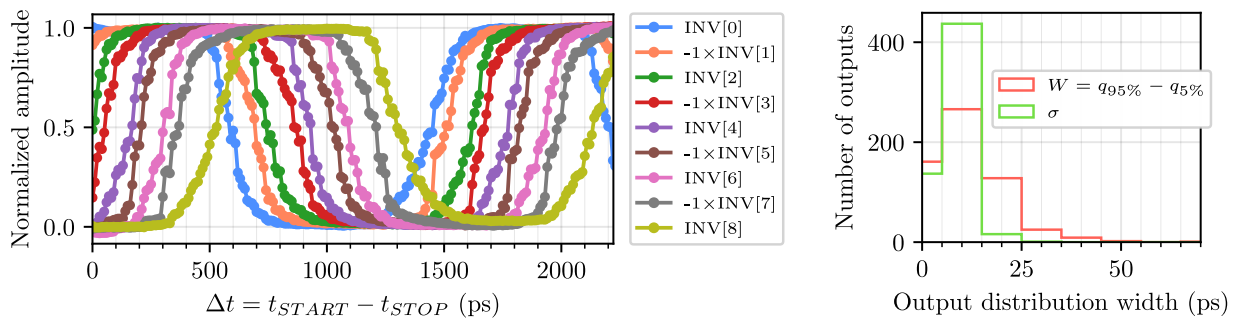

**Figure 6.** Left: Raw data from the analog INV[:] bus measured from the semi-analog TDC. Right: Distribution of the time spread of each output, measured with the "full width" $W$ and with the standard deviation $\sigma$, when digitized with a 3 bit ADC.

## 5. Conclusions

Two TDC designs were successfully implemented and tested with a focus on the constraints and requirements of 4D pixels for HEP applications, specifically for a timing layer replacement of disks in the TEPX detector. One of the designs successfully accomplished the required time resolution ($\sigma \sim \mathcal{O}(10 \text{ ps})$) while at the same time occupying the space requirements to fit within a $100 \times 100$ µm$^2$ pixel. Power consumption and radiation hardness studies are planned for the future.

There is ongoing work on an analog front end, which has already been produced in the same technology as the fully digital TDC, to characterize it and couple it with the presented TDC designs to simulate the full readout circuit. A timing readout chip prototype is planned to interface with an array of $30 \times 30$ LGAD pixels of $100 \times 100$ µm$^2$ area, which is currently in production.

# References

[1] CERN Document Server. "High-Luminosity Large Hadron Collider (HL-LHC): Technical Design Report V. 0.1." CERN, 2017. https://doi.org/10.23731/CYRM-2017-004.

[2] CMS, Collaboration. "A MIP Timing Detector for the CMS Phase-2 Upgrade." CERN Document Server, March 15, 2019. https://cds.cern.ch/record/2667167.

[3] CERN Document Server. "Technical Design Report: A High-Granularity Timing Detector for the ATLAS Phase-II Upgrade," June 5, 2020. https://cds.cern.ch/record/2719855.

[4] CERN Document Server. "The Phase-2 Upgrade of the CMS Tracker," June 30, 2017. https://cds.cern.ch/record/2272264.

[5] Currás, E., et al. "Inverse Low Gain Avalanche Detectors (ILGADs) for Precise Tracking and Timing Applications." Nuclear Instruments and Methods in Physics Research Section A: Accelerators, Spectrometers, Detectors and Associated Equipment 958 (April 2020): 162545. https://doi.org/10.1016/j.nima.2019.162545.

[6] Paternoster, G., et al. "Trench-Isolated Low Gain Avalanche Diodes (TI-LGADs)." IEEE Electron Device Letters 41, no. 6 (June 2020): 884–87. https://doi.org/10.1109/LED.2020.2991351.

[7] Tornago, M., et al. "Resistive AC-Coupled Silicon Detectors: Principles of Operation and First Results from a Combined Analysis of Beam Test and Laser Data." ArXiv:2007.09528 [Physics], October 16, 2020. http://arxiv.org/abs/2007.09528.

[8] Paternoster, G., et al. "Novel Strategies for Fine-Segmented Low Gain Avalanche Diodes." Nuclear Instruments and Methods in Physics Research Section A: Accelerators, Spectrometers, Detectors and Associated Equipment 987 (January 21, 2021): 164840. https://doi.org/10.1016/j.nima.2020.164840.

[9] Ayyoub, S., et al. "A New Approach to Achieving High Granularity for Silicon Diode Detectors with Impact Ionization Gain." ArXiv:2101.00511 [Physics], January 2, 2021. http://arxiv.org/abs/2101.00511.

[10] Cartiglia, N., et al. "LGAD Designs for Future Particle Trackers." Nuclear Instruments and Methods in Physics Research Section A: Accelerators, Spectrometers, Detectors and Associated Equipment 979 (November 1, 2020): 164383. https://doi.org/10.1016/j.nima.2020.164383.

[11] Characterization of delay in PSI test setup, Matías Senger, March 18 2021. Published in https://msenger.web.cern.ch/characterization-of-delay-in-psi-test-setup/.

[12] "Fully digital" TDC characterization, Matías Senger, May 11 2021. Published in https://msenger.web.cern.ch/fully-digital-tdc-characterization/.

[13] Development of a timing chip prototype in 110 nm CMOS technology, L. Caminada, B. Kilminster, A. Macchiolo, B. Meier, M. Senger, S. Wiederkehr. TIPP 2021 conference oral presentation, material available in https://sengerm.github.io/TIPP2021/.